\documentclass[conference]{IEEEtran}
\IEEEoverridecommandlockouts
\usepackage{cite}
\usepackage{url}
\usepackage{amsmath,amssymb,amsfonts}
\usepackage{algorithmic}
\usepackage{graphicx}
\usepackage{textcomp}
\usepackage{xcolor}
\usepackage{booktabs}
\usepackage{multirow}
\usepackage{bbding}
\usepackage{enumitem}
\usepackage[caption=false]{subfig}
\usepackage{graphicx}
\def\BibTeX{{\rm B\kern-.05em{\sc i\kern-.025em b}\kern-.08em
    T\kern-.1667em\lower.7ex\hbox{E}\kern-.125emX}}
\begin{document}

\title{Robust Generative Audio Quality Assessment: Disentangling Quality from Spurious Correlations}

% \author{Anonymous ICME submission}
\author{\IEEEauthorblockN{
Kuan-Tang Huang\IEEEauthorrefmark{1},
Chien-Chun Wang\IEEEauthorrefmark{1}\IEEEauthorrefmark{3},
Cheng-Yeh Yang\IEEEauthorrefmark{1},
Hung-Shin Lee\IEEEauthorrefmark{4},
Hsin-Min Wang\IEEEauthorrefmark{2},
and Berlin Chen\IEEEauthorrefmark{1}
}
\IEEEauthorblockA{\IEEEauthorrefmark{1}Dept. Computer Science and Information Engineering, National Taiwan Normal University, Taiwan}
\IEEEauthorblockA{\IEEEauthorrefmark{2}Institute of Computer Science, Academia Sinica, Taiwan}
\IEEEauthorblockA{\IEEEauthorrefmark{3}E.SUN Financial Holding Co., Ltd., Taiwan}
\IEEEauthorblockA{\IEEEauthorrefmark{4}United Link Co., Ltd., Taiwan}
}
\maketitle

\begin{abstract}
The rapid proliferation of AI-Generated Content (AIGC) has necessitated robust metrics for perceptual quality assessment.
However, automatic Mean Opinion Score (MOS) prediction models are often compromised by data scarcity, predisposing them to learn spurious correlations---such as dataset-specific acoustic signatures---rather than generalized quality features.
To address this, we leverage domain adversarial training (DAT) to disentangle true quality perception from these nuisance factors.
Unlike prior works that rely on static domain priors, we systematically investigate domain definition strategies ranging from explicit metadata-driven labels to implicit data-driven clusters.
Our findings reveal that there is no "one-size-fits-all" domain definition; instead, the optimal strategy is highly dependent on the specific MOS aspect being evaluated.
Experimental results demonstrate that our aspect-specific domain strategy effectively mitigates acoustic biases, significantly improving correlation with human ratings and achieving superior generalization on unseen generative scenarios.
\end{abstract}


\section{Introduction}
The exponential growth of AI-Generated Content (AIGC) has revolutionized the multimedia landscape.
Particularly within the audio modality, generative audio has established itself as a cornerstone of modern content creation.
This domain encompasses a diverse array of tasks, ranging from text-to-speech (TTS) to text-to-music (TTM) and universal text-to-audio (TTA).
These advanced technologies are now driving a wide range of applications, such as creating dynamic soundscapes for immersive streaming, generating background scores for personalized media, and enabling interactive audio for virtual environments.
However, accurately evaluating the perceptual quality of these generated sounds remains a critical challenge.
While subjective listening tests yielding mean opinion scores (MOS) represent the gold standard for assessment, they are notoriously expensive and time-consuming to conduct.
Consequently, the development of automatic MOS prediction models has become indispensable, serving as a scalable and efficient proxy for human ratings \cite{saeki2022, deshmukh2024, zezario2023}.

However, the reliability of these data-driven models is often severely compromised by the lack of large-scale subjectively labeled data.
In such low-resource regimes, models are predisposed to learn spurious correlations rather than generalized quality features.
Within limited training sets, high subjective ratings may coincidentally align with specific, non-quality-related acoustic characteristics intrinsic to the source data.
For instance, a model might erroneously learn to associate high quality with the specific timbre of a musical instrument, the background ambiance of an environmental recording, or a particular room reverberation pattern in speech, simply because these traits dominate the highly rated samples in the limited training corpus.
Consequently, the model overfits to these nuisance factors---treating them as proxies for quality.
When deployed to unseen data where these specific acoustic signatures are absent, the model's predictions become unreliable, highlighting the need to disentangle true quality perception from these domain-specific biases.

The necessity of disentangling quality from confounding factors is echoed in adjacent fields.
In video quality assessment, \cite{wu2023} attempts to separate aesthetics from technical quality by engineering specific input views.
Similarly, in speaker recognition, recent works \cite{liu2023} have proposed specialized Gaussian inference layers to disentangle static speaker traits from dynamic linguistic content.
However, these approaches often rely on either intricate, hand-crafted heuristics or complex, task-specific architectural designs to define the separation boundaries.
To avoid reliance on rigid heuristics or task-specific designs, we introduce a generalized domain adversarial training (DAT) \cite{ganin2016} framework capable of automatically purging dataset-induced biases.
By leveraging this strategy, we enforce the model to discard these nuisance factors within the latent space, retaining only the features pertinent to intrinsic perceptual quality.

While DAT has been widely adopted in speech and audio tasks---such as addressing variations arising from different noise conditions \cite{shinohara2016}, accents \cite{sun2018}, or recording corpora \cite{abdelwahab2018, tanaka2022}---these works typically treat domain definition as a static prior.
In contrast, we argue that for MOS prediction under data scarcity, the optimal definition of a "domain" is not self-evident.
A critical yet underexplored question arises: what constitutes the most effective "domain" for adversarial training?
Crucially, our experiments suggest there is no "one-size-fits-all" answer; instead, we find that different MOS aspects necessitate specific domain definition strategies to maximize prediction generalization.

To this end, we systematically investigate three distinct domain definition strategies:
1) Source-based, which leverages explicit metadata (e.g., dataset identity) as static priors;
2) K-means clustering, which discovers implicit, data-driven acoustic patterns in the latent space, where we further examine the impact of cluster granularity ($K$) on adaptation performance; and
3) Random assignment, serving as a control baseline to validate the necessity of meaningful domain structures.
Our findings reveal that the choice of domain definition is pivotal.
By identifying the optimal adversarial target among these strategies, we effectively mitigate specific acoustic biases and achieve superior generalization across diverse, unseen generative scenarios.
The main contributions of this study\footnote{Our code: \url{https://github.com/610494/domainGRL}.} are summarized as follows:
\begin{itemize}[noitemsep,leftmargin=*]
    \item \textbf{Addressing Spurious Correlations:} We identify data scarcity causes overfitting to acoustic signatures, proposing a DAT framework to mitigate this without complex heuristics.
    \item \textbf{Systematic Investigation of Domain Definitions:} We systematically explore effective adversarial targets, ranging from explicit metadata to implicit data-driven clusters.
    \item \textbf{Generalizability:} We demonstrate that our findings on domain granularity are robust across various backbone models.
\end{itemize}

% ==========================================
% Section 3: Methodology
% ==========================================
\section{Proposed Method}
\label{sec:method}

We propose a robust MOS prediction framework that incorporates Domain Adversarial Training (DAT) to learn quality-aware representations invariant to domain shifts.
In this section, we detail the model architecture and systematically investigate domain definition strategies.

\subsection{Model Architecture}
\label{sec:model_architecture}
\begin{figure}[t]
\centering
% User requested 0.8 width previously to reduce size
\includegraphics[width=0.85\linewidth]{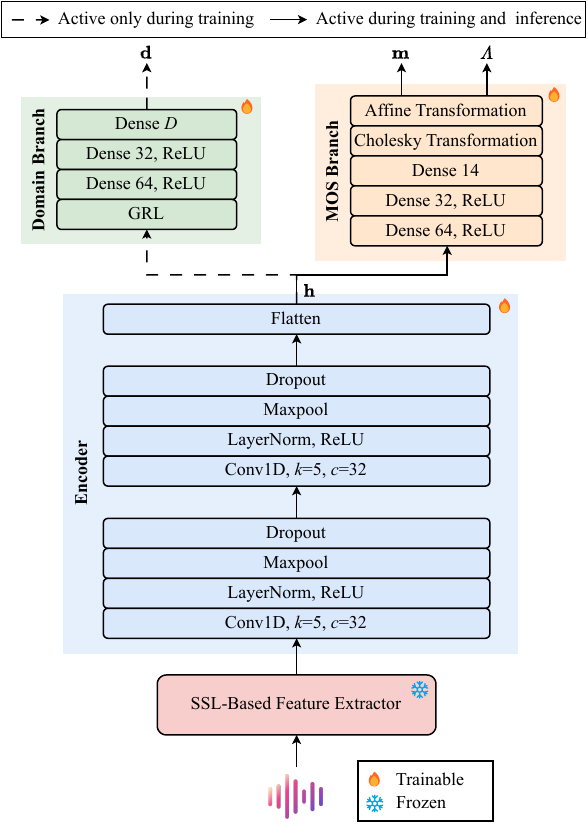}
\vspace{-10pt}
\caption{
The proposed model architecture with DAT. 
% The framework incorporates a Domain Discriminator connected via a GRL to enforce domain invariance.
% \textbf{Optimization Strategy:} As annotated in the figure, the upstream SSL feature extractor is frozen to preserve general acoustic knowledge, while the downstream Encoder and MOS branches are fully trainable to adapt to the quality assessment task.
}
\label{fig:main}
\vspace{-15pt}
\end{figure}

The overall framework is illustrated in Fig. \ref{fig:main}.
The model comprises three key components: a pre-trained SSL feature extractor, a quality prediction backbone (i.e., MultiGauss), and a domain adversarial branch.

\textbf{SSL-Based Feature Extractor:} 
Given the vast diversity of pre-trained SSL models, we specifically select the XLS-R 2B model \cite{babu2022} due to its exceptional model capacity and training scale. 
While it is primarily trained on speech, previous works \cite{ragano2023, wu2022} show that speech-based SSL representations can accurately assess the quality of vinyl music collections and encode general audio events, such as environmental sounds, with high fidelity. 
By leveraging this broad acoustic knowledge, we utilize XLS-R as a general-purpose encoder to ensure stable quality assessment across the diverse speech, music, and audio conditions in our study.

\textbf{MultiGauss Backbone (MOS Predictor):}
We leverage the state-of-the-art MultiGauss framework \cite{cumlin2025} as our backbone.
It extracts a flattened latent representation $\mathbf{h}$ to predict a multivariate mean vector $\mathbf{m}$ (representing quality scores across multiple aspects) and a covariance matrix $\mathbf{\Lambda}$.
The matrix $\mathbf{\Lambda}$ explicitly models predictive uncertainty and captures correlations between these quality dimensions.
While $\mathbf{\Lambda}$ models predictive uncertainty, our analysis focuses on the latent representation $\mathbf{h}$ and the mean vector $\mathbf{m}$ to align directly with MOS-based evaluation metrics.
This ensures that the domain adaptation process prioritizes the most salient features responsible for quality score prediction.

\textbf{Domain Discriminator:}
To disentangle domain-specific information, we introduce a parallel ``Domain Branch'' connected to the shared representation $\mathbf{h}$ via a Gradient Reversal Layer (GRL) \cite{ganin2015}.
This branch comprises stacked dense layers culminating in an output layer of size $D$ (where $D$ denotes the number of domains), mapping the features to a domain prediction vector $\mathbf{d}$.
During training, the GRL reverses the gradients flowing from this discriminator to the encoder, effectively forcing $\mathbf{h}$ to become invariant to the domain distinctions.

\textbf{Optimization Objective:}
The entire framework is trained end-to-end using a multi-task objective:
\begin{equation}
    \mathcal{L}_{total} = \mathcal{L}_{task} + \lambda \mathcal{L}_{adv}
\end{equation}

Following MultiGauss \cite{cumlin2025}, we employ the Gaussian Negative Log-Likelihood (GNLL) loss as $\mathcal{L}_{task}$ to estimate the multivariate parameters ($\mathbf{m}$ and $\mathbf{\Lambda}$).
$\mathcal{L}_{adv}$ denotes the standard cross-entropy loss for domain classification.
Through the GRL, the gradients from $\mathcal{L}_{adv}$ are reversed during backpropagation, thereby enforcing the encoder to learn domain-invariant representations while minimizing the prediction error.

\subsection{Domain Definition Strategies}
\label{sec:domain_definitions}

The effectiveness of the GRL hinges on the definition of the adversarial target.
Unlike prior works that rely on fixed domain labels, we systematically explore three strategies covering explicit, latent, and stochastic definitions:

\begin{itemize}[noitemsep,leftmargin=*]
    \item \textbf{DAT-Source (Dataset Origin):} 
    This explicit strategy utilizes the inherent dataset identifiers (e.g., AudioSet vs. LibriTTS) as domain labels ($N=6$). It aims to capture macro-level variations in production environments, such as differences in recording equipment, codec standards, and post-processing pipelines specific to each data source.
    
    \item \textbf{DAT-Kmeans (Latent Acoustic):} 
    To capture acoustic variations that transcend dataset boundaries, we employ unsupervised K-means clustering on the pre-trained acoustic embeddings extracted from the training set. 
    Specifically, we utilize the last-layer representations from the same frozen SSL backbone used for MOS prediction, applying temporal mean pooling to obtain global utterance-level embeddings for standard K-means clustering (using Euclidean distance).
    We treat the number of clusters $K$ as a dynamic hyperparameter representing the granularity of the domain definition.
    We explore a range of granularities (e.g., $K \in \{2, \dots, 10\}$), selected to encompass the number of explicit source datasets ($N=6$), to identify the optimal resolution for capturing fine-grained, implicit acoustic textures---such as reverberation patterns or background noise profiles---that are often not annotated but significantly impact the domain distribution.

    \item \textbf{DAT-Random (Perturbation):} 
    This strategy assigns random labels to samples.
    It serves as a baseline to verify whether performance gains stem from meaningful domain disentanglement or merely from the stochastic regularization effect of the gradient reversal mechanism.
\end{itemize}

% \subsection{Optimization Objective}
% \label{sec:objective}
% The entire framework is trained end-to-end using a multi-task objective defined as:
% \begin{equation}
%     \mathcal{L}_{total} = \mathcal{L}_{task} + \lambda \mathcal{L}_{adv}
% \end{equation}
% where $\mathcal{L}_{task}$ is the regression loss for MOS prediction, and $\mathcal{L}_{adv}$ is the domain classification loss. 
% The GRL reverses the gradients from $\mathcal{L}_{adv}$ during backpropagation, forcing the encoder to learn domain-invariant features while minimizing prediction error.

\section{Experimental Setup}
\label{sec:experimental_setup}

\subsection{Dataset}
% We evaluate our proposed method on the \textbf{AES-Natural dataset} \cite{tjandra2025}, a comprehensive corpus specifically curated to benchmark audio quality assessment across heterogeneous acoustic domains.
% The dataset consists of 2,950 audio clips stratified into three primary categories: \textit{Speech} (950 clips sourced from EARS, LibriTTS, and Common Voice), \textit{Music} (1,000 clips from MUSDB18 and MusicCaps), and \textit{General Environmental Audio} (1,000 clips from AudioSet).
% To ensure reliable ground truth, all samples were annotated by a panel of 10 expert listeners possessing professional backgrounds in audio engineering or music theory.
We evaluate our proposed method on the AES-Natural dataset \cite{tjandra2025}, utilizing a rigorous split protocol to benchmark generalization from natural acoustic priors to unseen generative scenarios.
The Training and Validation sets consist of natural recordings stratified into three categories: Speech (EARS, LibriTTS, and Common Voice), Music (MUSDB18 and MusicCaps), and General Audio (AudioSet).
Due to source availability, the final training set comprises 2,544 clips (approx. 31.6 hours), and the validation set consists of 232 clips (approx. 3.2 hours).
In contrast, the Evaluation set is strictly disjoint, containing 3,060 machine-generated audio samples (approx. 7.9 hours) synthesized by various generative models.
To ensure reliable ground truth, all samples were annotated by a panel of 10 expert listeners possessing professional backgrounds in audio engineering or music theory.

Departing from traditional MOS datasets that strictly evaluate technical degradation, AES-Natural characterizes audio perception along four distinct dimensions.
This multi-dimensional schema allows us to disentangle low-level signal fidelity from inherent content attributes:

\begin{itemize}[noitemsep,leftmargin=*]
    \item \textbf{Production Quality (PQ):} Reflects the \textit{low-level technical fidelity} of the signal. This metric focuses on physical signal degradations, such as noise floor, distortion, and bandwidth limitations, which are typically dependent on the recording equipment and channel characteristics.
    
    \item \textbf{Production Complexity (PC):} Quantifies the structural richness and density of the audio content (e.g., the number of active stems in a mix or the layering of sound effects). While strongly correlated with content type, simple models may risk forming spurious correlations between this metric and specific dataset signatures rather than the content itself.
    
    \item \textbf{Content Enjoyment (CE):} Represents the \textit{intrinsic aesthetic appeal} and engagement value of the audio.
    As an intrinsic aesthetic attribute, CE abstracts away from simple signal fidelity but can be susceptible to bias arising from listener preferences for specific genres or recording styles.
    
    \item \textbf{Content Usefulness (CU):} Assesses the functional utility of the audio for its intended application (e.g., speech intelligibility or atmospheric immersion for environmental sounds).
\end{itemize}

\subsection{Training Setup}
\label{sec:training_setup}

To verify generalizability, we integrate the DAT strategy into two distinct backbone architectures.
First, for \textbf{MultiGauss \cite{cumlin2025}}, we follow the original implementation, training for 30 epochs with a batch size of 64 and a learning rate of $1\times 10^{-4}$.
The best checkpoint is selected based on the lowest validation loss.
Second, we evaluate \textbf{Audiobox-Aesthetics \cite{tjandra2025}}, an architecture that directly predicts quality from multi-layer WavLM features without an additional encoder.
Unlike the frozen backbone in MultiGauss, we fully fine-tune this encoder for 200 epochs (batch size 16, learning rate $1\times 10^{-5}$) to enable adversarial gradient propagation.
Both models are optimized using AdamW with 0 weight decay.

The adversarial loss weight $\mathbf{\Lambda}$ acts as a hyperparameter controlling the trade-off between task performance and domain invariance.
Through empirical validation, we set $\lambda=0.5$ for the DAT-Source strategy to effectively bridge the significant distribution gaps between distinct datasets.
Conversely, for the DAT-Kmeans and DAT-Random strategies, we set $\lambda=0.1$.
This reduced weight is crucial for the DAT-Kmeans strategy to prevent over-regularization since the implicit clusters may partially encode quality-related information that should not be aggressively suppressed.
For the DAT-Kmeans strategy, we specifically set $K=8$ as the default granularity, which will be further analyzed in Sec. \ref{sec:impact_of_k}.

\begin{table*}[t]
\centering
\caption{Performance comparison with existing baselines across four aspects: Production Quality (PQ), Production Complexity (PC), Content Enjoyment (CE), and Content Usefulness (CU).
The symbol $\dagger$ indicates results cited from original papers. Best performance in each column is highlighted in \textbf{bold}.}
\label{tab:domain_comparison}
\setlength{\tabcolsep}{12pt} % 欄位減少，可稍微加大間距提升閱讀性
\begin{tabular}{l cc cc cc cc}
\toprule
\multirow{2}{*}{\textbf{System / Strategy}} & \multicolumn{2}{c}{\textbf{PQ (Technical)}} & \multicolumn{2}{c}{\textbf{PC (Content)}} & \multicolumn{2}{c}{\textbf{CE (Content)}} & \multicolumn{2}{c}{\textbf{CU (Functional)}} \\
\cmidrule(lr){2-3} \cmidrule(lr){4-5} \cmidrule(lr){6-7} \cmidrule(lr){8-9}
 & \textbf{MSE}$\downarrow$ & \textbf{SRCC}$\uparrow$ & \textbf{MSE}$\downarrow$ & \textbf{SRCC}$\uparrow$ & \textbf{MSE}$\downarrow$ & \textbf{SRCC}$\uparrow$ & \textbf{MSE}$\downarrow$ & \textbf{SRCC}$\uparrow$ \\
\midrule

% --- 區塊 1: Existing Baselines ---
\multicolumn{9}{l}{\textit{\textbf{Existing Baselines (SOTA)}}} \\
QAMRO$^{\dagger}$ \cite{wang2025} & - & 0.883 & - & 0.942 & - & 0.869 & - & 0.852 \\
DRASP$^{\dagger}$ \cite{yang2025} & - & 0.900 & - & 0.936 & - & 0.890 & - & 0.911 \\
AESA-Net$^{\dagger}$ \cite{wisnu2025} & 0.635 & 0.896 & \textbf{0.198} & 0.928 & 3.991 & 0.904 & \textbf{0.533} & 0.894 \\
\midrule
MultiGauss \cite{cumlin2025} & 0.557 & 0.942 & 1.093 & 0.947 & 1.841 & 0.938 & 0.945 & 0.961 \\
\hspace{3mm} + L2 Regularization \cite{loshchilov2019} & 0.472 & 0.941 & 0.962 & 0.944 & 1.634 & 0.945 & 0.874 & 0.962 \\
\hspace{3mm} + High Dropout & 0.649 & 0.944 & 1.894 & 0.945 & 2.182 & 0.965 & 1.060 & 0.947 \\
\hspace{3mm} + \textbf{DAT-Source} (Ours) & 0.413 & 0.940 & 0.747 & \textbf{0.969} & \textbf{1.581} & \textbf{0.967} & 0.855 & 0.959 \\
\hspace{3mm} + \textbf{DAT-Kmeans} (Ours) & 0.479 & \textbf{0.953} & 0.928 & 0.945 & 1.605 & 0.952 & 0.835 & \textbf{0.963} \\
\hspace{3mm} + \textbf{DAT-Random} (Ours) & \textbf{0.390} & 0.941 & 0.945 & 0.958 & 1.689 & 0.961 & 0.789 & 0.959 \\
\bottomrule
\vspace{-15pt}
\end{tabular}
\end{table*}

\subsection{Evaluation Metrics}
\label{sec:evaluation_metrics}
To comprehensively assess our model, we report system-level Mean Squared Error (MSE) and Spearman’s Rank Correlation Coefficient (SRCC).
Following the evaluation protocols of prominent MOS prediction challenges \cite{huang2025}, all metrics are calculated by first averaging the predictions and ground-truth labels for all utterances belonging to the same generative system.
SRCC is prioritized as the primary metric to reflect the model's capability to reliably rank diverse generative systems.
By combining it with system-level MSE, which assesses absolute precision and model calibration, we ensure a robust evaluation that disentangles ranking consistency from absolute scale errors in cross-domain scenarios.

\section{Results and Discussion}
\label{sec:results}

\begin{figure}[t]
    \centering
    \includegraphics[width=1.0\linewidth]{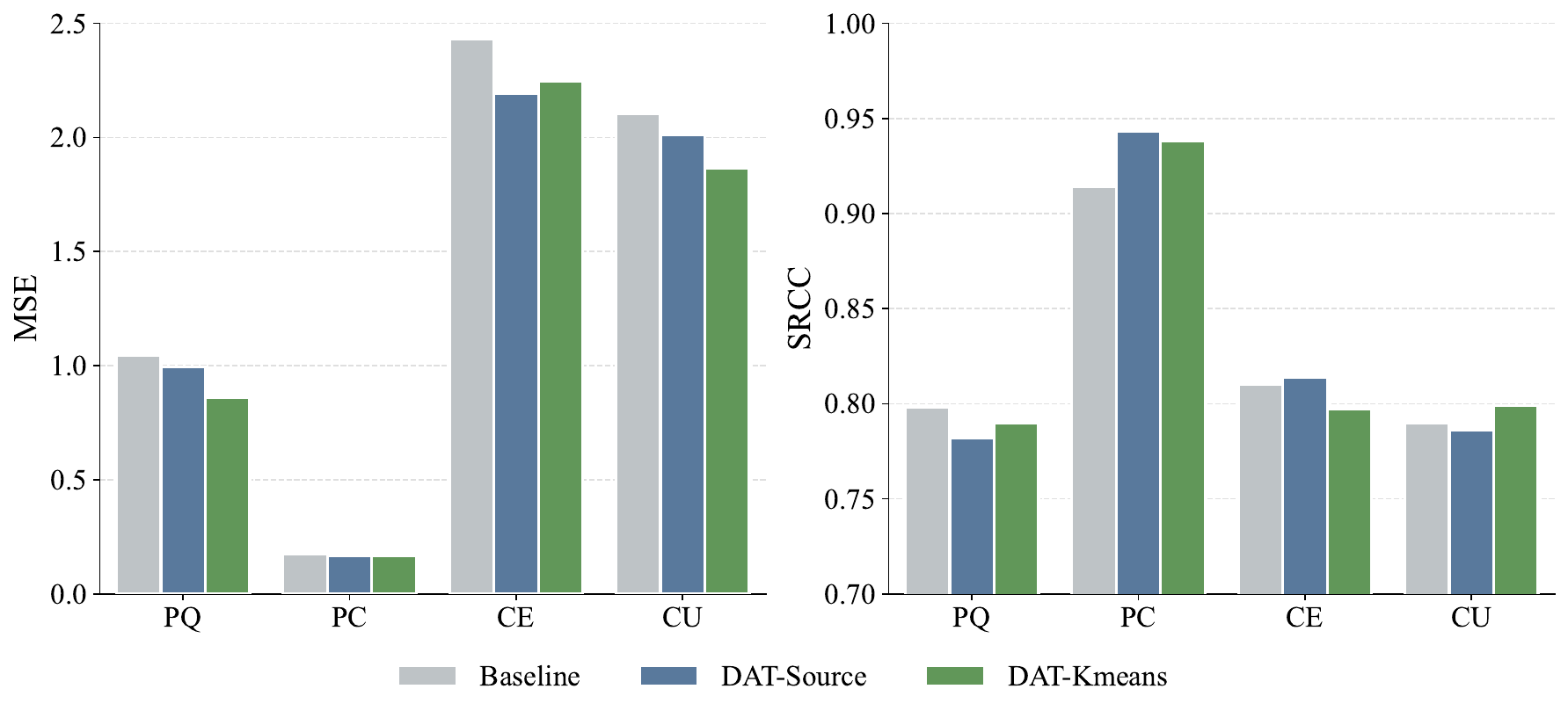}
    \vspace{-20pt}
    \caption{Performance comparison on Audiobox-Aesthetics across MSE and SRCC.
    The results are reported for four aspects: PQ, PC, CE, and CU.}
    \label{fig:backbone_b_performance}
    \vspace{-20pt}
\end{figure}

\subsection{Main results}
\label{sec:main_results}
We evaluate the effectiveness of the proposed domain adversarial training (DAT) framework on the MultiGauss backbone.
Table \ref{tab:domain_comparison} presents the performance comparison. 
The results demonstrate that explicitly disentangling domain information consistently improves model robustness.
Following \cite{cooper2022}, a two-sided t-test ($p \le 0.05$) confirms that the performance gains of our proposed DAT strategies are statistically significant compared to the baseline.

Our experiments reveal that the optimal definition of a ``domain'' is inherently dimension-dependent, reflecting the distinct nature of different perceptual attributes.

For inherent content attributes, specifically Production Complexity (PC) and Content Enjoyment (CE), the DAT-Source strategy yields the most significant improvements.
Since these attributes exhibit systematic biases---for instance, music datasets inherently yield much higher complexity scores than speech datasets---the baseline model is prone to ``shortcut learning'' based on dataset signatures.
DAT-Source penalizes this reliance, significantly reducing PC MSE from 1.093 to 0.747 while achieving the highest SRCC of 0.969.
This improvement stems from forcing the encoder to learn intrinsic structural representations rather than relying on source identity.

In contrast, technical and functional attributes, such as Production Quality (PQ) and Content Usefulness (CU), achieve optimal performance under the DAT-Kmeans strategy.
Unlike content-related biases, technical degradations (e.g., background noise, reverberation) frequently transcend dataset boundaries and overlap across sources.
We observe an interesting trade-off here: while explicit source labels help calibrate the absolute score scale (reducing PQ MSE to 0.413), the latent acoustic clusters discovered by K-means better capture fine-grained texture variations essential for preserving relative rankings.
This is evidenced by DAT-Kmeans achieving the highest SRCC of 0.953 for PQ.
Thus, for technical metrics where domain distributions overlap, unsupervised acoustic clustering offers a superior adversarial target for refining ranking capabilities.

To verify whether performance gains stem from blind regularization rather than principled domain invariance, we compared our strategies against L2 regularization, High Dropout, and DAT-Random.
While L2 regularization and DAT-Random provide improvements in absolute error (MSE) for certain aspects, they consistently fail to match the superior ranking performance of our aspect-specific DAT strategies in terms of SRCC.
Crucially, SRCC serves as our primary evaluation metric as it directly reflects the model's capability to reliably rank generative systems, a task where traditional and stochastic regularization prove inadequate compared to our targeted disentanglement approach.
Furthermore, increasing stochasticity via High Dropout leads to statistically significant performance degradation.
Notably, our framework consistently outperforms these traditional and stochastic regularization techniques across the majority of dimensions. These results confirm that targeted domain disentanglement is fundamentally superior to blind generic regularization.
By explicitly addressing the specific nature of each quality dimension, our framework effectively purges spurious correlations and empowers state-of-the-art backbones to capture more intrinsic and generalized quality features.

\textit{Generalization across Model Architectures.}
To verify whether the effectiveness of our aspect-specific strategies generalizes across different frameworks, we evaluate the proposed method on the Audiobox-Aesthetics \cite{tjandra2025} architecture.
Unlike the MultiGauss backbone, which uses frozen XLS-R features, this model utilizes fine-tuned WavLM representations, providing a distinct feature space for validation.
As illustrated in Fig. \ref{fig:backbone_b_performance}, the performance trends are highly consistent with previous observations: the DAT-Source strategy remains superior for inherent content attributes (PC, CE), while DAT-Kmeans consistently excels in technical and functional dimensions (PQ, CU).
This alignment across different backbone architectures and SSL feature extractors validates the robustness of our domain definition strategies and confirms their benefits are independent of the underlying model configuration.

\begin{figure}[t]
    \centering
    \includegraphics[width=0.9\linewidth]{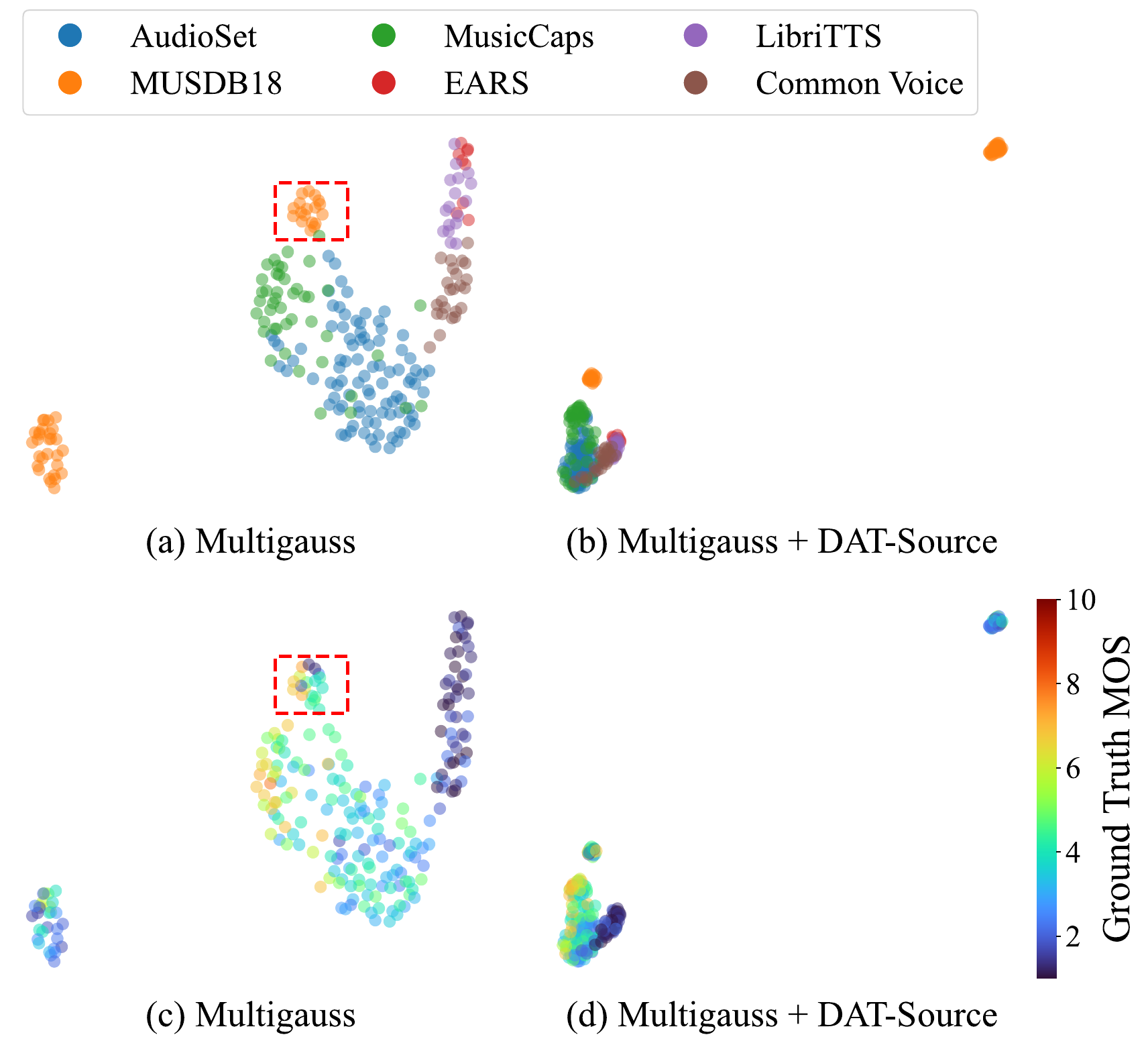}
    % \vspace{-10pt}
    \caption{Visualization of the $\mathbf{h}$ on the development set using UMAP. 
    The top row is colored by source domain labels, and the bottom by PC scores.
    }
    \vspace{-15pt}
    \label{fig:umap_2d}
\end{figure}

\subsection{Latent Space Analysis}
\label{sec:latent_space_analysis}

To analyze the manifold structure and verify whether the DAT framework effectively removes domain-specific information from the latent space, we project the bottleneck features $h$ obtained from the model encoder into a two-dimensional space using UMAP. 
As illustrated in Fig. \ref{fig:umap_2d} (top row), the baseline model exhibits severe domain bias.
Specifically, the region highlighted by the red dashed box in Fig. \ref{fig:umap_2d}(a) shows a tight cluster formed solely by dataset identity.
However, referencing Fig. \ref{fig:umap_2d}(c), it becomes evident that samples within this domain-driven cluster possess vastly different quality scores.
This clustering fragments the semantic space: high-quality samples are isolated within their respective domain ``islands'' rather than forming a cohesive high-quality region.
This confirms that the baseline learns spurious correlations, grouping samples by domain signatures rather than their actual perceptual quality.
In contrast, our proposed method successfully merges these heterogeneous domains into a unified manifold, indicating the removal of non-causal signatures.
Crucially, this alignment preserves intrinsic quality information, as Fig. \ref{fig:umap_2d}(d) reveals a continuous quality gradient transitioning from low to high quality across the aligned manifold.

\begin{figure}[t]
    \centering
    \subfloat[Multigauss]{
        \includegraphics[width=0.46\linewidth]{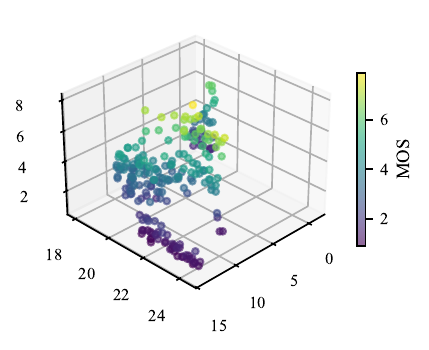}
    }
    \hfill
    \subfloat[Multigauss + DAT-Source]{
        \includegraphics[width=0.46\linewidth]{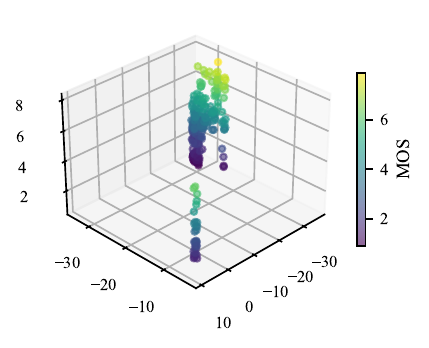}
    }
    
    % \caption{3D visualization of the ``Quality Terrain'' generated by combining 2D UMAP projections of encoder features $\mathbf{h}$ with the predicted MOS as the z-axis. 
    % (a) The baseline features form separated clusters, indicating that domain signatures interfere with the continuity of the quality latent space. 
    % (b) Our proposed method aligns heterogeneous domains into a cohesive ``Quality Pillar,'' confirming that achieving domain invariance preserves and even clarifies the intrinsic quality ranking.}
    \caption{3D ``Quality Terrain'' generated by combining 2D UMAP projections of encoder features $\mathbf{h}$ with the predicted MOS as the z-axis for (a) the baseline and (b) our proposed DAT strategy.}
    \vspace{-15pt}
    \label{fig:umap_3d}
\end{figure}

To further investigate the relationship between domain invariance and quality prediction, we extend the UMAP projection to three dimensions by incorporating the predicted MOS as the vertical axis, creating a ``Quality Terrain'' visualization. This allows us to inspect whether the latent manifold maintains a consistent structural hierarchy across different score ranges. 

As shown in Fig. \ref{fig:umap_3d}(a), the baseline model's features remain scattered into fragmented clusters across the 3D space.
Even at identical quality levels, samples from different domains are horizontally segregated, forcing the model to navigate a disjointed latent space.
In contrast, our DAT strategy (Fig. \ref{fig:umap_3d}(b)) collapses these horizontal domain variances into a cohesive ``Quality Pillar.''
In this structure, the manifold aligns vertically according to the quality gradient, where samples from all heterogeneous domains are successfully mapped onto a shared, continuous trajectory.
This vertical alignment confirms that our domain-adversarial objective does not compromise the ranking capability; instead, it enforces a principled representation where domain-invariant features and quality-relevant information are effectively disentangled and organized.

% To mathematically quantify domain disentanglement, we evaluate linear probing performance on the frozen latent features $\mathbf{h}$ (Table \ref{tab:linear_probing}).
% \textbf{(1) Domain Disentanglement:} The baseline's high Domain Acc. (90.9\%) proves heavy entanglement with dataset origins.
% \textbf{DAT-Source} degrades this predictability (87.5\%), proving the successful purging of spurious signatures.
% Conversely, \textbf{DAT-Kmeans} increases Acc. (92.2\%) because clustering by acoustic textures inadvertently sharpens dataset-specific recording conditions, mathematically validating our dimension-dependent theory.
% \textbf{(2) Technical Attributes (PQ):} Echoing Table \ref{tab:domain_comparison}, because DAT-Kmeans effectively organizes these acoustic textures, it yields the optimal linear manifold for predicting PQ (0.800).
% \textbf{(3) Content Attributes (PC):} The baseline achieves an artificially high PC SRCC (0.891) by exploiting dataset identities as a shortcut.
% DAT-Source purposefully dismantles this cheating shortcut (dropping to 0.879), which precisely explains why it generalizes significantly better in unseen zero-shot scenarios (Table \ref{tab:domain_comparison}).
Linear probing on $\mathbf{h}$ (Table \ref{tab:linear_probing}) reveals the baseline's severe dataset entanglement (90.9\% Domain Acc.) artificially inflates its PC SRCC (0.891) via identity shortcuts. DAT-Source effectively purges these spurious signatures (87.5\% Acc.), intentionally dismantling this shortcut (PC drops to 0.879) for superior zero-shot generalization (Table \ref{tab:domain_comparison}). Conversely, DAT-Kmeans inadvertently increases predictability (92.2\%) by clustering acoustic textures; this effectively organizes the linear manifold, yielding the optimal predictor for technical attributes like PQ (0.800).

\subsection{Impact of Domain Granularity and Grouping Strategy}
\label{sec:impact_of_k}

\begin{figure}[t]
    \centering
    \includegraphics[width=\linewidth]{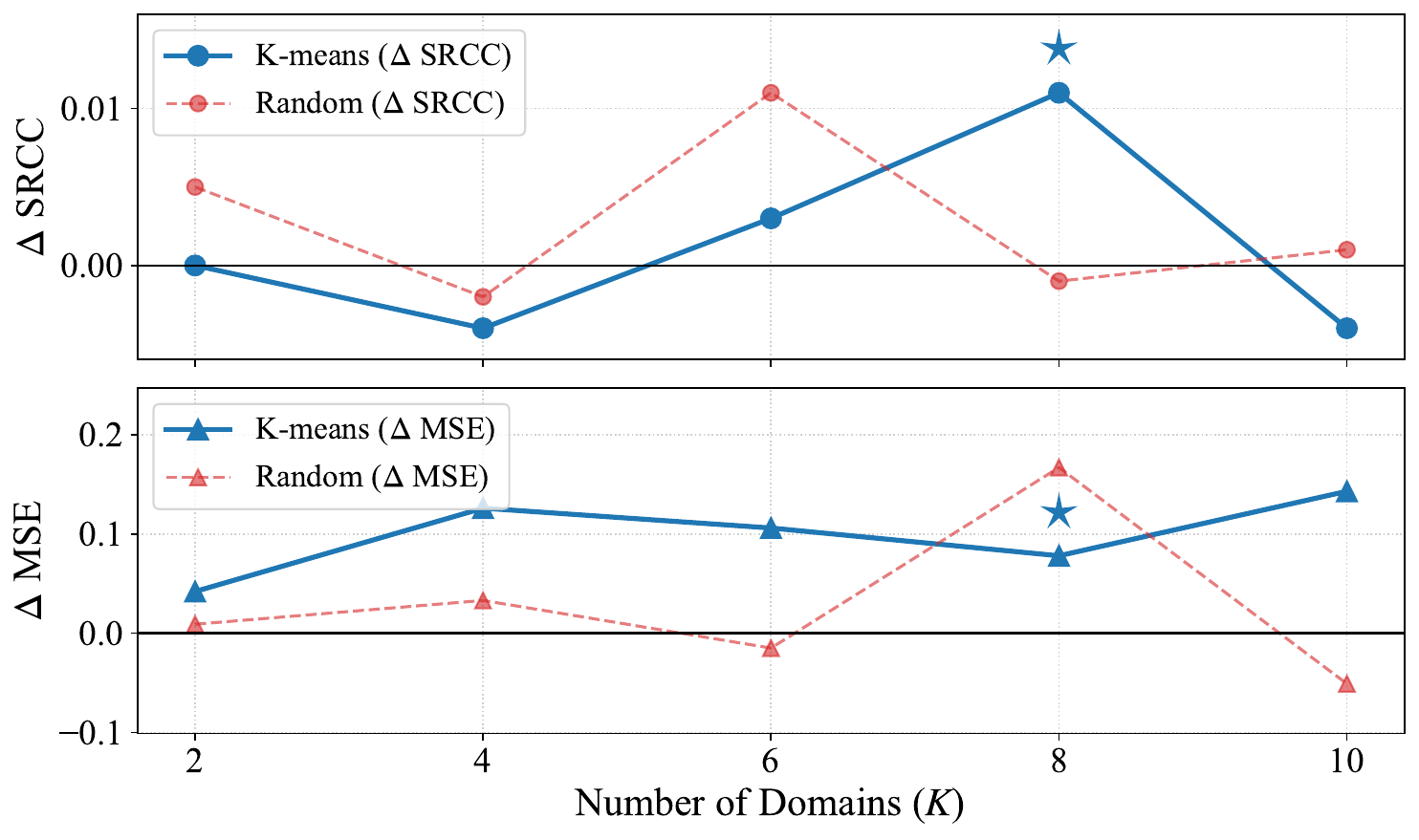}
    \vspace{-20pt}
    \caption{Ablation study on domain granularity $K$ for the PQ dimension.
    The top panel shows the absolute improvement in SRCC ($\Delta$ SRCC), and the bottom panel shows the improvement in MSE ($\Delta$ MSE) relative to the baseline.
    The star ($\star$) denotes the optimal configuration at $K=8$, which yields the most balanced gains across both metrics.}
    \label{fig:ablation_k}
    \vspace{-15pt}
\end{figure}

To investigate the impact of domain granularity $K$, we evaluated model performance across $K \in \{2, 4, 6, 8, 10\}$, centered around $K=6$ to provide a direct comparison with the DAT-Source strategy.
This analysis aims to determine whether latent acoustic clusters offer more precise domain disentanglement than explicit dataset identities. As illustrated in Fig. \ref{fig:ablation_k}, the proposed DAT-Kmeans strategy demonstrates a more structured and superior performance trend compared to the random assignment baseline.
The DAT-Kmeans strategy reaches its performance peak at $K=8$ (marked with $\star$ in Fig. \ref{fig:ablation_k}), achieving the highest gain in ranking consistency ($\Delta\text{SRCC} \approx 0.011$) and a significant reduction in error ($\Delta\text{MSE} \approx 0.08$).
While $K=10$ yields a slightly higher MSE gain, its $\Delta\text{SRCC}$ drops below the baseline, suggesting that over-partitioning the acoustic space introduces noise that hinders the model's ranking ability.

In contrast, the random strategy exhibits high instability. Although it shows sporadic gains in MSE (e.g., at $K=8$), its impact on SRCC is erratic and often falls into the negative range, indicating a degradation in ranking capability compared to the baseline.
This disparity confirms that domain definitions must be anchored in meaningful acoustic sub-structures to provide reliable adversarial gradients.
The superior trajectory of K-means validates our hypothesis that data-driven clustering effectively captures the underlying domain bias essential for robust audio quality assessment.

\begin{table}[t]
\centering
\caption{Linear probing analysis on latent features $\mathbf{h}$.}
\label{tab:linear_probing}
\resizebox{\columnwidth}{!}{
\begin{tabular}{@{}lccc@{}}
\toprule
Strategy & Domain Acc. (\%, $\downarrow$) & PQ SRCC ($\uparrow$) & PC SRCC ($\uparrow$) \\ \midrule
Baseline & 90.9 & 0.795 & \textbf{0.891} \\
\textbf{DAT-Source} & \textbf{87.5} & 0.798 & 0.879 \\
\textbf{DAT-Kmeans} & 92.2 & \textbf{0.800} & 0.886 \\ \bottomrule
\end{tabular}
}
\vspace{-15pt}
\end{table}

\section{Conclusion and Future Work}

In this paper, we introduced Domain Adversarial Training (DAT) to the task of MOS prediction to address the critical issue of shortcut learning. Specifically, we proposed an \textit{aspect-specific} DAT framework, demonstrating that by forcing the encoder to be invariant to domain factors, we can significantly improve the robustness and generalization of quality assessment models. Our analysis reveals that the optimal definition of a ``domain'' is inherently dimension-dependent: explicit source labels are superior for disentangling content-related biases, while latent acoustic clusters are more effective for refining technical quality rankings. 

These strategies consistently empower state-of-the-art backbones to capture intrinsic quality features rather than spurious correlations. Future work will focus on developing a unified multi-branch architecture that simultaneously integrates both explicit source constraints and latent acoustic clustering. By leveraging these complementary domain definitions, we aim to build a robust, universal model that achieves optimal performance across all perceptual dimensions of audio quality.

\bibliographystyle{IEEEbib}
\bibliography{references}

@inproceedings{loshchilov2019,
	title = {Decoupled weight decay regularization},
	booktitle = {Proc. {ICLR}},
	author = {Loshchilov, Ilya and Hutter, Frank},
	year = {2019},
}

@inproceedings{huang2025,
	title = {The {audioMOS} challenge 2025},
	booktitle = {Proc. {ASRU}},
	author = {Huang, Wen-Chin and Wang, Hui and Liu, Cheng and Wu, Yi-Chiao and Tjandra, Andros and Hsu, Wei-Ning and Cooper, Erica and Qin, Yong and Toda, Tomoki},
	year = {2025},
}

@inproceedings{ganin2015,
	title = {Unsupervised domain adaptation by backpropagation},
	booktitle = {Proc. {ICML}},
	author = {Ganin, Yaroslav and Lempitsky, Victor},
	year = {2015},
}

@inproceedings{wu2022,
	title = {The efficacy of self-supervised speech models for audio representations},
	booktitle = {Proc. {HEAR}},
	author = {Wu, Tung-Yu and Hsu, Tsu-Yuan and Li, Chen-An and Lin, Tzu-Han and Lee, Hung-yi},
	year = {2022},
}

@inproceedings{ragano2023,
	title = {Audio quality assessment of vinyl music collections using self-supervised learning},
	booktitle = {Proc. {ICASSP}},
	author = {Ragano, Alessandro and Benetos, Emmanouil and Hines, Andrew},
	year = {2023},
}

@inproceedings{babu2022,
	title = {{XLS}-{R}: {Self}-supervised cross-lingual speech representation learning at scale},
	booktitle = {Proc. {Interspeech}},
	author = {Babu, Arun and Wang, Changhan and Tjandra, Andros and Lakhotia, Kushal and Xu, Qiantong and Goyal, Naman and Singh, Kritika and von Platen, Patrick and Saraf, Yatharth and Pino, Juan and Baevski, Alexei and Conneau, Alexis and Auli, Michael},
	year = {2022},
}

@inproceedings{liu2023,
	title = {Disentangling voice and content with self-supervision for speaker recognition},
	booktitle = {Proc. {NeurIPS}},
	author = {Liu, Tianchi and Lee, Kong Aik and Wang, Qiongqiong and Li, Haizhou},
	year = {2023},
}

@article{abdelwahab2018,
	title = {Domain adversarial for acoustic emotion recognition},
	volume = {26},
	journal = {IEEE/ACM Transactions on Audio, Speech, and Language Processing},
	author = {Abdelwahab, Mohammed and Busso, Carlos},
	year = {2018},
	pages = {2423--2435},
}

@inproceedings{wu2023,
	title = {Exploring video quality assessment on user generated contents from aesthetic and technical perspectives},
	booktitle = {Proc. {ICCV}},
	author = {Wu, Haoning and Zhang, Erli and Liao, Liang and Chen, Chaofeng and Hou, Jingwen and Wang, Annan and Sun, Wenxiu and Yan, Qiong and Lin, Weisi},
	year = {2023},
}

@inproceedings{deshmukh2024,
	title = {{PAM}: {Prompting} audio-language models for audio quality assessment},
	booktitle = {Proc. {Interspeech}},
	author = {Deshmukh, Soham and Alharthi, Dareen and Elizalde, Benjamin and Gamper, Hannes and Al Ismail, Mahmoud and Singh, Rita and Raj, Bhiksha and Wang, Huaming},
	year = {2024},
}

@inproceedings{saeki2022,
	title = {{UTMOS}: {UTokyo}-{SaruLab} system for {VoiceMOS} challenge 2022},
	booktitle = {Proc. {Interspeech}},
	author = {Saeki, Takaaki and Xin, Detai and Nakata, Wataru and Koriyama, Tomoki and Takamichi, Shinnosuke and Saruwatari, Hiroshi},
	year = {2022},
}

@inproceedings{wang2025,
	title = {{QAMRO}: {Quality}-aware adaptive margin ranking optimization for human-aligned assessment of audio generation systems},
	booktitle = {Proc. {ASRU}},
	author = {Wang, Chien-Chun and Huang, Kuan-Tang and Yang, Cheng-Yeh and Lee, Hung-Shin and Wang, Hsin-Min and Chen, Berlin},
	year = {2025},
}

@inproceedings{wisnu2025,
	title = {Improving perceptual audio aesthetic assessment via triplet loss and self-supervised embeddings},
	booktitle = {Proc. {ASRU}},
	author = {Wisnu, Dyah A. M. G. and Zezario, Ryandhimas E. and Rini, Stefano and Wang, Hsin-Min and Tsao, Yu},
	year = {2025},
}

@inproceedings{yang2025,
	title = {{DRASP}: {A} dual-resolution attentive statistics pooling framework for automatic {MOS} prediction},
	booktitle = {Proc. {APSIPA} {ASC}},
	author = {Yang, Cheng-Yeh and Huang, Kuan-Tang and Wang, Chien-Chun and Lee, Hung-Shin and Wang, Hsin-Min and Chen, Berlin},
	year = {2025},
}

@article{zezario2023,
	title = {Deep learning-based non-intrusive multi-objective speech assessment model with cross-domain features},
	volume = {31},
	journal = {IEEE/ACM Transactions on Audio, Speech, and Language Processing},
	author = {Zezario, Ryandhimas E. and Fu, Szu-Wei and Chen, Fei and Fuh, Chiou-Shann and Wang, Hsin-Min and Tsao, Yu},
	year = {2023},
	pages = {54--70},
}

@inproceedings{tjandra2025,
	title = {Meta audiobox aesthetics: {Unified} automatic quality assessment for speech, music, and sound},
	booktitle = {Arxiv preprint {arXiv}:2502.05139},
	author = {Tjandra, Andros and Wu, Yi-Chiao and Guo, Baishan and Hoffman, John and Ellis, Brian and Vyas, Apoorv and Shi, Bowen and Chen, Sanyuan and Le, Matt and Zacharov, Nick and Wood, Carleigh and Lee, Ann and Hsu, Wei-Ning},
	year = {2025},
}

@inproceedings{cooper2022,
	title = {Generalization ability of {MOS} prediction networks},
	booktitle = {Proc. {ICASSP}},
	author = {Cooper, Erica and Huang, Wen-Chin and Toda, Tomoki and Yamagishi, Junichi},
	year = {2022},
}

@inproceedings{sun2018,
	title = {Domain adversarial training for accented speech recognition},
	booktitle = {Proc. {ICASSP}},
	author = {Sun, Sining and Yeh, Ching-Feng and Hwang, Mei-Yuh and Ostendorf, Mari and Xie, Lei},
	year = {2018},
}

@inproceedings{shinohara2016,
	title = {Adversarial multi-task learning of deep neural networks for robust speech recognition},
	booktitle = {Proc. {Interspeech}},
	author = {Shinohara, Yusuke},
	year = {2016},
}

@article{ganin2016,
	title = {Domain-adversarial training of neural networks},
	volume = {17},
	number = {59},
	journal = {Journal of Machine Learning Research},
	author = {Ganin, Yaroslav and Ustinova, Evgeniya and Ajakan, Hana and Germain, Pascal and Larochelle, Hugo and Laviolette, François and March, Mario and Lempitsky, Victor},
	year = {2016},
	pages = {1--35},
}

@inproceedings{tanaka2022,
	title = {Domain adversarial self-supervised speech representation learning for improving unknown domain downstream tasks},
	booktitle = {Proc. {Interspeech}},
	author = {Tanaka, Tomohiro and Masumura, Ryo and Sato, Hiroshi and Ihori, Mana and Matsuura, Kohei and Ashihara, Takanori and Moriya, Takafumi},
	year = {2022},
}

@inproceedings{cumlin2025,
	title = {Multivariate probabilistic assessment of speech quality},
	booktitle = {Proc. {Interspeech}},
	author = {Cumlin, Fredrik and Liang, Xinyu and Ungureanu, Victor and Reddy, Chandan K. A. and Schüldt, Christian and Chatterjee, Saikat},
	year = {2025},
}

\end{document}